\documentclass{mnras}
\usepackage[T1]{fontenc}
\usepackage{ae,aecompl}
\usepackage{graphicx}	% Including figure files
\usepackage{amsmath}	% Advanced maths commands
\usepackage{amssymb}	% Extra maths symbols
\DeclareGraphicsExtensions{.ps}

\title{SuperModel Predictions in the Outskirts of the Galaxy Cluster Zwicky 3146}
\author[R.Fusco-Femiano]{
Roberto Fusco-Femiano$^{1}$\thanks{E-mail: roberto.fusco@inaf.it}
\\
$^{1}$IAPS-INAF, via del Fosso del Cavaliere, 00133 Roma, Italy}

\date{Accepted XXX. Received YYY; in original form ZZZ}
\pubyear{2015}

\begin{document}
\label{firstpage}
\pagerange{\pageref{firstpage}--\pageref{lastpage}}
\maketitle

\begin{abstract}
The relaxed galaxy cluster Zwicky 3146 is analyzed via the SuperModel, a tool already tested on many clusters since 2009. In particular, this analysis is focused on the intracluster medium X-ray temperature data measured by \textit{XMM-Newton} up to $r_{500}$. A previous analysis was based on the temperature profile derived from the Sunyaev-Zeld$^{\prime}$ovich effect pressure data.  
The gas mass fraction $f_{gas}$ is obtained from the resulting SuperModel temperature profile extrapolated up to the virial radius $R$, that turns out in agreement with the steep temperature profiles observed by \textit{Suzaku}, and from the gas density profile observed by \textit{XMM-Newton}. The comparison between $f_{gas}$ with 
the universal value indicates a non-thermal pressure component, $p_{nth}$, in the cluster outskirts. The SuperModel analysis shows a ratio $\alpha(R)$ ($\simeq 50\%$) of $p_{nth}$ to the total pressure greater than the values found by simulations, highlighting the possible presence of accreting  substructures and inhomogeneities in the gas density profile. Once that this profile is corrected for clumpiness, the level of $p_{nth}$ is considerably reduced. However, a significant turbulence ($\alpha(R)\simeq 20\%$) and entropy flattening are still present in the outskirts of the galaxy cluster Zwicky 3146.
\end{abstract}

\section{Introduction}
 
The sloshing cool core cluster Zwicky 3146 at z = 0.291 has been recently observed at 90 GHz by MUSTANG-2 (Romero et al. 2019; hereafter R19). 
The pressure profile derived by the Sunyaev-Zeld$^{\prime}$ovich (SZ) effect (1972) is in excellent agreement with that derived in X rays by \textit{XMM-Newton} observations. From the SZ pressure profile R19 derive 
$M_{500}$ and $M_{2500}$ via three methods: by means of a $Y - M$ scaling relation ($Y$ is a volumetric integral of the thermal 
electron pressure), the second employs the hydrostatic equilibrium (HE) assuming spherical symmetry, and the third considers the virial theorem. The electron density profile is inferred from the \textit{XMM-Newton} data. The authors consider a non-parametric (NP) pressure model, where a power-law distribution of pressure is assumed within twelve radial bins between $5^{\prime\prime}$ and $5^{\prime}$, and the fit with
the generalized NFW pressure profile (Nagai et al. 2007) fixing some parameters. The estimates of
$M_{500}$ depend on the model and estimation method. With the NP model R19 derive $8.16^{+0.44}_{-0.54}\times 10^{14} M_{\odot}$ from a $Y-M$ relation, $8.29^{+1.93}_{-1.24}\times 10^{14} M_{\odot}$ from hydrostatic
equilibrium, and $9.05^{+0.56}_{-0.67}\times 10^{14} M_{\odot}$ from the virial equilibrium; for the gNFW model the values are $7.70\pm 0.17\times 10^{14} M_{\odot}$, $8.33^{+1.93}_{-1.24}\times 10^{14} M_{\odot}$, and $10.60\pm 0.10\times 10^{14} M_{\odot}$, respectively.

R19 extend their analysis to other thermodynamic quantities such as gas temperature and entropy. The temperature profile is obtained by combining the thermal electron pressure $P_e$ given by SZ data with the electron density $n_e$,
through the relation $k_B T_e = P_e/n_e$. The above quantities define the entropy parameter $K_e = k_B T_e/n_e^{2/3} =  P_e/n_e^{5/3}$ as reported in Voit (2005).

Here, it is exploited the capability of the SuperModel (SM; Cavaliere, Lapi \& Fusco-Femiano 2009) to extrapolate the intracluster medium (ICM) temperature profile, observed by \textit{XMM-Newton} up to $\sim r_{500}$, to the virial radius assumed to be $R = 2 r_{500}$. 
Since 2009, the SM has been used to investigate the ICM thermodynamic properties of several clusters (Fusco-Femiano, Cavaliere \& Lapi 2009; Fusco-Femiano et al. 2011). This semi-analytic tool is based on few physical parameters that define the entropy state of the intracluster medium. 
%The SM is a semi-analytic tool to investigate the ICM thermodynamic properties
%based on few physical parameters of the underlying entropy state of the intracluster medium. Since 2009, several clusters have been successfully analyzed with the SuperModel (Fusco-Femiano, Cavaliere \& Lapi 2009; Fusco-Femiano et al. 2011). 
As shown in Sect. 2, the SM temperature profile includes the possible presence of a non-thermal pressure component to
sustain the hydrostatic equilibrium. This has allowed more recently to highlight the role of the nonthermal pressure component in the cluster outskirts using the knowledge of the universal gas mass fraction (Fusco-Femiano \& Lapi 2013, 2014, 2015, 2018; Fusco-Femiano 2019). 
This method has been used by Eckert et al (2019) to constrain the level of the non-thermal support in the galaxy clusters of the
XMM Cluster Outskirts Project (X-COP) (Eckert et al. 2017).

According to hydrodynamic simulations the level of the non-thermal pressure
component, generated by bulk and turbulent motions during the formation of large-scale structures, increases going toward the outskirts of galaxy clusters (Vazza et al. 2009; Valdarnini 2011; Gaspari \& Churasov 2013; Lau et al. 2013; Nelson et al. 2014). The contribution of these non-thermal processes to the overall pressure results to be $\sim 30\%$ in the cluster outskirts. Instead, in cluster cores this contribution seems to amount to only some per cent as reported by the \textit{Hitomi} observations of the Perseus cluster (the turbulent pressure is 4\% or less of the thermodynamic pressure, Hitomi collaboration 2016).  
Determining this level is fundamental for understanding the ICM thermodynamic properties. In fact, these peripherals regions, connecting the ICM with the surrounding environment, are sites of physical processes and events (see Kravtsov \& Borgani 2012; Cavaliere \& Lapi 2013; Reiprich er al. 2013; Walker et al. 2019). Besides, the total cluster mass is biased low when the non-thermal pressure component is omitted in the HE. Non-thermal energy is present also in relaxed clusters as shown by the SM analysis of several relaxed clusters (see the references above).

Several relaxed clusters observed by \textit{Suzaku} in their outskirts show features like: steep decline of the ICM temperature in the region $r\sim (0.3-1)r_{200}$ (see Akamatsu et al 2011; Reiprich et al. 2013, Walker et al. 2013), 
flattening of the entropy profile at $r\gtrsim 0.5r_{200}$ (see Walker et al. 2012; 2013) relative to the shape $k\propto r^{1.1}$ expected under pure gravitational infall (see Tozzi \& Norman 2011; Lapi et al. 2005; Voit 2005), significant azimuthal variations of the ICM thermodynamic properties (see Kawaharada et al. 2010; Ichikawa et al. 2013; Sato et al. 2014), an unphysical decreasing behavior of the total mass at large radii (Kawaharada et al. 2010; Walker et al. 2012; Ichikawa et al. 2013; Sato et al. 2014).
 
The scope of this paper is to investigate through the SM analysis of the X-ray temperature data, measured by \textit{XMM-Newton} up to $\sim r_{500}$, whether some of the features reported by \textit{Suzaku} can be recovered in the outskirts of the relaxed cluster Zwicky 3146. In particular, to check if also for this cluster the \textit{XMM-Newton} temperature data are consistent, as for the X-COP cluster sample, with the rapid decline of the temperature observed by \textit{Suzaku} in the outskirts of several clusters. One of the conclusions discussed in Fusco-Femiano (2019) is that a steep temperature profile could be present in the outskirts of the X-COP cluster sample instead of the flatter temperature profile derived by the SZ observations reported in Ghirardini et al. (2019). A further objective is to investigate the possible
presence of a non-thermal component in the cluster outskirts that is fundamental to determine for the reasons above reported. 

These goals are pursued exploiting the capability of the SuperModel to extrapolate the temperature profile out to the virial radius. 
This capability has been successfully tested in previous SM analyses. For Abell 2142 (Fusco-Femiano \& Lapi 2018) the derived hydrostatic mass profile is consistent with all the 
measurements reported in the literature not only at $r_{500}$ but also at greater distances, in particular with the $M_{200}$ value obtained
from Subaru weak lensing (Umetsu et al. 2009). Recently, Fusco-Femiano (2019) has shown that the SM temperature profile derived from the stacked temperature profile of the X-COP cluster sample (Ghirardini et al. 2018), observed by \textit{XMM-Newton} up to $r_{500}$, is consistent with the temperature profiles observed by \textit{Suzaku} in several cluster outskirts. In addition, as shown in Sect. 3, the SM temperature 
extrapolation for Zwicky 3146 is in agreement with the temperature profile given by the gNFW pressure model based on SZ observations (Romero et al. 2019).

As reported above, the SM analysis includes the possibility to determine the level of a possible non-thermal component in the outskirts of Zwicky 3146, not reported from the analysis of R19. When applied to the X-COP cluster sample the SM analysis derives a ratio
of the non-thermal pressure component to the total one ($\alpha = p_{nth}/p_{tot}$) in agreement with the median values derived by Eckert et al. (2019) at $r_{500}$ ($\alpha \sim 6\%$) and $r_{200}$ ($\alpha \sim 10\%$) when the fit is performed to the joint X-ray and SZ temperature profiles. Greater values of $\alpha$ are instead obtained when the fit is to the X-ray data only, in agreement with the numerical simulations of Nelson et al. (2014) and Martizzi \& Agrusa (2016). At the virial radius $\alpha (R)$ is in the range (20-40)\%.

The paper is organized as follows. In the next Section 2, the temperature profile derived by $\textit{XMM-Newton}$ observations of the galaxy cluster Zwicky 3146 is analyzed with the SuperModel. Section 3 reports 
the gas mass fraction, the entropy and the total mass profiles obtained by combining the SM temperature and the $\textit{XMM-Newton}$ electron density profiles. In particular, from the gas mass fraction $f_{gas}$ it is possible to infer the level of the non-thermal pressure component present in the cluster outskirts. The results and the conclusions are drawn in Section 4. 
%6The SM equations are reported in Appendix A, yielding the temperature, pressure and total mass also when a nonthermal pressure component is included in the HE. The temperature, pressure and entropy profiles are normalized to their median values at $r_{500}$ reported in Section 2.

Throughout the paper the standard flat cosmology is adopted with parameters in
round numbers: $H_0 = 70$ km s$^{-1}$ Mpc$^{-1}$, $\Omega_{\Lambda} = 0.7$,
$\Omega_M = 0.3$ (\textit{Planck} collaboration XIII 2016).

With this cosmology, R19 report at $z=0.291$ $r_{2500}=130^{\prime\prime}$, $r_{500}=293^{\prime\prime}$ yielding a virial radius $R=2r_{500}=586^{\prime\prime}$ assumed in this paper; one arcsecond corresponds to 4.36 kpc.

\section{SuperModel analysis of the galaxy cluster Zwicky 3146}

% The X-ray electron density profile $n_e$ has been corrected for the presence of clumpiness applying the azimuthal median method reported in Eckert et al. (2015). The presence of inhomogeneities in the accreted gas is expected to be relevant for the X-ray flux measured beyond $r_{500}$ (see Vazza et al 2013; Roncarelli et al 2013;
%Eckert et al. 2015). This method allows to
%disentangle the effects of gas clumping from the possible presence of a nonthermal pressure component in the outskirts of galaxy clusters.

The X-ray temperature profile of the galaxy cluster Zwicky 3146 has been obtained from the $\textit{XMM-Newton}$ derived pressure and electron density profiles reported in Romero et al. (2019) (see Fig.1). 
The analysis of the temperature profile is performed with the entropy-based SM derived by the HE equation when the entropy distribution $k = k_B T/n^{2/3}$ is specified. The assumed entropy profile starts with a central entropy value $k_c$ given by feedback from astrophysical sources and radiative cooling; then it increases with a power law with slope $a$ up to the virial radius $R$ where the entropy $k_R$ is produced by supersonic gas inflows from the surrounding environment into the  dark matter (DM) gravitational potential well (see Cavaliere et al. 2009 for more details). This entropy shape is represented by $k(r) = k_c + (k_R - k_c)(r/R)^a$ (Voit 2005).

%and then increases  is $k(r) = k_c + (k_R - k_c)(r/R)^a$ (Voit 2005) where $k_c$ is the central entropy level set by feedback from astrophysical sources and radiative cooling; $k_R$ is the entropy at the virial radius $R$ produced by supersonic gas inflows from the surrounding environment into the DM gravitational potential well; $a$ is the slope of the power law increase from $k_c$ (see Cavaliere et al. 2009 for more details).
 
The HE equation yields the SM temperature profile

\begin{equation*}
\frac{T(r)}{T_R} = \left[\frac{k(r)}{k_R}\right]^{3/5}\, \left[\frac{1 +
\delta_R}{1 + \delta(r)}\right]^{2/5}\,\times
\end{equation*}
\begin{equation}
\left\{1 + \frac{2}{5}\frac{b_R}{1 +
\delta_R}\int_r^R {\frac{{\rm d}x}{x} \frac{v^2_c(x)}{v^2_R}\,
\left[\frac{k_R}{k(x)}\right]^{3/5}\, \left[\frac{1 + \delta_R}{1 +
\delta(x)}\right]^{3/5}}\right\}
\end{equation}

\par\noindent
where $v_c$ is the DM circular velocity ($v_R$ is the value at the virial radius $R$), and $b_R$
is the ratio at $R$ of $v^2_c$ to the sound speed squared (Cavaliere et al. 2009; Cavaliere et al. 2011). This temperature profile includes  the possible presence of a non-thermal pressure component, $p_{nth}$,
that added to the thermal one, $p_{th}$, gives a total pressure $p_{tot}(r) = p_{th}(r) + p_{nth}(r) = p_{th}(r)\left[ 1 +\delta(r)\right]$ where $\delta(r) = p_{nth}/p_{th}$. The functional shape $\delta(r)$ is given by $\delta(r) = \delta_R\, e^{-(R-r)^2/l^2}$ in agreement with the indication of numerical simulations (e.g., Lau et al. 2009; Vazza et al. 2011). $\delta(r)$ decays on the scale $l$ toward the inside (see Cavaliere, Lapi \& Fusco-Femiano 2011). 

A modified entropy shape for the SM has been taken into consideration by Lapi, Fusco-Femiano \& Cavaliere (2010) to satisfy the rapid decline 
of the temperature and the
entropy flattening observed by \textit{Suzaku} in the outskirts of several cool-core clusters (Akamatsu et al. 2011; Reiprich et al. 2013; Walker et al. 2013), and in the directions of non-cool core ones not disturbed by mergers (like Coma, Simionescu et al. 2013). This entropy profile
starts as a power law with slope $a$, but it has a linear entropy decline with gradient $a^{\prime}\equiv (a - a_R)/(R/r_b -1)$ at distances greater than $r_b$; the free parameters $r_b$ and $a^{\prime}$ are determined from the fitting of the temperature profile.
In the SM fitting procedure the model parameters reported in Table I can assume any value.
%To satisfy the steep temperature and flat entropy profiles observed by \textit{Suzaku} toward the virial radius in several CC clusters (Akamatsu et al. 2011; Reiprich et al. 2013; Walker et al. 2013), and in the directions of NCC ones not disturbed by mergers (like Coma, Simionescu et al. 2013), Lapi, Fusco-Femiano \& Cavaliere (2010) have considered a modified entropy shape for the SM that starts as a power law with slope $a$, but flattens at distances greater than $r_b$. For the sake of simplicity, a linear entropy decline with gradient $a^{\prime}\equiv (a - a_R)/(R/r_b -1)$ is assumed, where $r_b$ and $a^{\prime}$ are free parameters to be determined from the fitting of the temperature profile. 

The SM fit shown in Fig. 1 (red dashed area) clearly indicates, although the $\textit{XMM-Newton}$ temperature data are limited to $\sim r_{500}$, a steep decline of the temperature at $r\gtrsim r_b$. The fit is obtained assuming a flattening of the entropy distribution in the cluster outskirts, in agreement with the \textit{Suzaku} observations reported above. Conversely, the entropy shape that increases with a power law with slope $a$ up to the virial radius ($a^{\prime}=0$) is inadequate to fit the temperature points (see blue curve of Fig. 1). A particular attention has been addressed to
the last point at $\sim r_{500}$ (green point) that may appear as biased low. Omitting this point the SM analysis gives a value of the free parameter $k_BT_R$ (temperature at the virial radius) only 5.7\% greater than that reported in Table I ($\delta_R=0$). This implies negligible differences in the results from the two analyses. A comparison between the SM temperature profile and those obtained by R19 using the NP and gNFW pressure profiles is shown in Fig. 2. The NP temperature profile declines more rapidly than the SM profile in the cluster outskirts. Instead, the SM and gNFW temperature profiles are absolutely consistent. 
\begin{figure}
\includegraphics[width=\columnwidth]{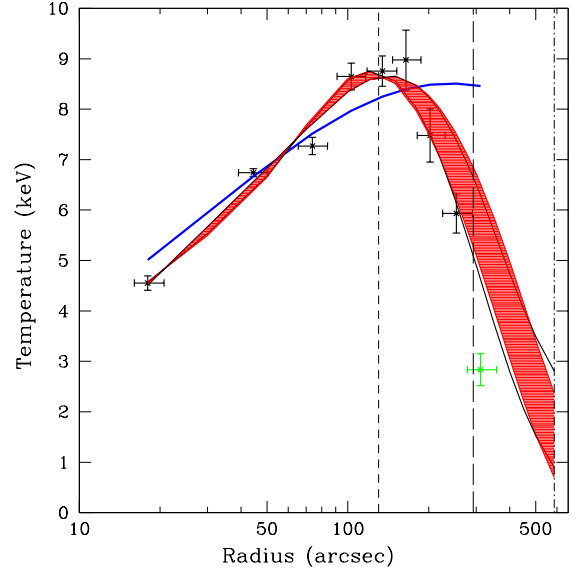}
\caption{Black points
represent the X-ray temperature data measured by \textit{XMM-Newton} (Romero et al. 2019). For the green point see the text. The blue line is the SM fit with $a^{\prime}=0$, while the dashed area is with $a^{\prime}>0$ (for both the fits $\delta_R=0$). The black curves represent the fit with $a^{\prime}>0$ and $\delta_R=1$. The vertical dashed, long-dashed and dot-dashed lines represent $r_{2500}$, $r_{500}$, and the virial radius $R$, respectively.}
\label{fig:zw3146_temp_figure}
\end{figure}

\begin{figure}
\includegraphics[width=\columnwidth]{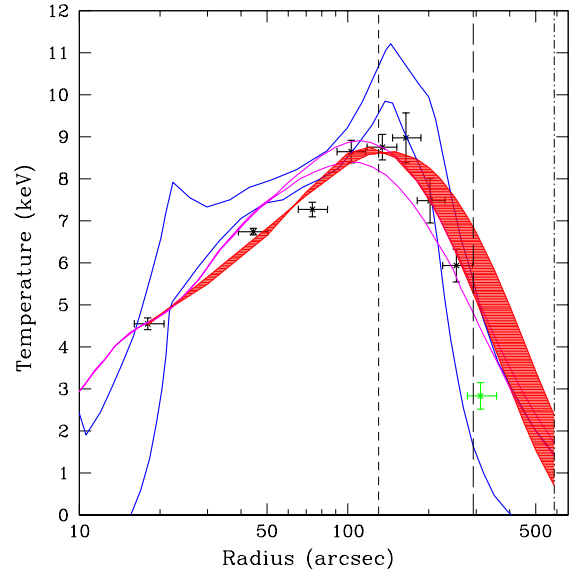}
\caption{Black points, green point and dashed area as in Fig. 1. The blue curves represent the temperature profile derived from the NP model,
while the magenta curves are derived from the gNFW model (Romero et al. 2019).
The vertical dashed, long-dashed and dot-dashed lines represent $r_{2500}$, $r_{500}$, and the virial radius $R$, respectively.}
\label{fig:zw3146_temp_np_nfw_figure}
\end{figure}

\begin{figure}
\includegraphics[width=\columnwidth]{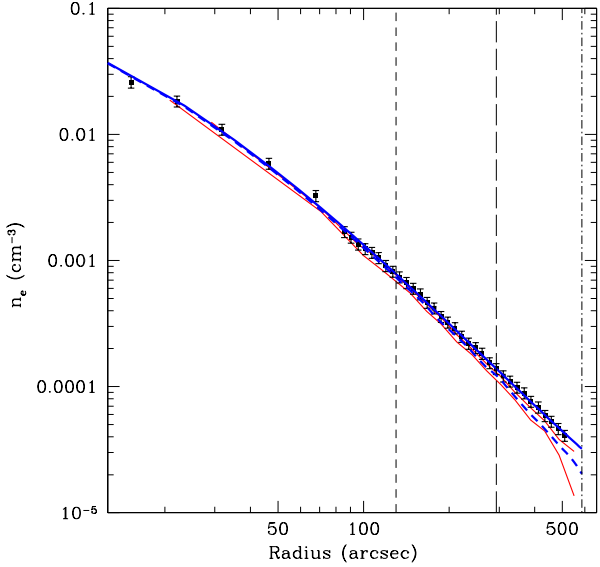}
\caption{The density points are derived from the \textit{XMM-Newton} observations. In particular, the density points at $r\lesssim 80^{\prime\prime}$ are derived from the NP pressure and temperature profiles (see Fig.s 4 and 7 in R19, respectively) through the relation $n_e=P_{NP}/k_BT_{NP}$, while the density points at greater distances are given by Fig. 12 (R19). The continuous curve is
obtained from the fit to the density points using the SM density profile reported in Cavaliere et al. (2009). The dashed curve represents the SM fit after that the median profile of the clumping factor
$\sqrt(C)$, derived by Nagai \& Lau (2011) for clusters with $M_{200}>10^{14} h^{-1} M_{\odot}$, is applied to the continuous density profile; the red lines represent the $1\sigma$ scatter of the clumping factor.  
The vertical dashed, long-dashed and dot-dashed lines represent $r_{2500}$, $r_{500}$, and the virial radius $R$, respectively.}
\label{fig:zw3146_dens_figure}
\end{figure}

\section{Non-thermal pressure component in the outskirts of Zwicky 3146}

The presence of a non-thermal pressure component in the outskirts of galaxy clusters can be established by comparing the cluster gas mass fraction with the universal baryon fraction (Planck collaboration XIII 2016) corrected for the baryon fraction in the form of stars (Gonzalez, Zaritsky \& Zabludoff 2007).
The matter content of galaxy clusters is expected to be approximately the universal value. 

The gas mass fraction $f_{gas} = M_{gas}/M_{tot}$ (see Eq. A1 to derive the gas mass $M_{gas}$ and the total cluster mass $M_{tot}$) is built by combining the SM temperature profile with the electron density data given by \textit{XMM-Newton} observations.
The SM temperature profile (see the red dashed area of Fig. 1) is obtained from the SM fit to the $\textit{XMM-Newton}$ temperature points when only the thermal pressure contributes to the cluster HE ($\delta_R = 0$ in Eq.1). The electron density profile (see the continuous curve in Fig. 3) is derived by fitting the density points with the SM density profile derived by Cavaliere et al. (2009).
Fig. 12 of R19 provides density points only at $r\gtrsim 80^{\prime\prime}$; while to have density points at lower distances it is necessary to
resort to the relation $n_e = P_{NP}/k_BT_{NP}$ where the fitted pressure $P_{NP}$ and the gas temperature $T_{NP}$ profiles are shown in Fig.s 4 and 7 of R19, respectively. The resulting SM $f_{gas}$ profile in Fig. 4 is above the universal value going toward the virial radius, consistent with the unphysical decline of the total mass reported in the same figure. This result highlights the presence of a non-thermal pressure support in the cluster outskirts that must be added to the thermal one in the HE equation to obtain the hydrostatic equilibrium.
 
New fits have been performed to the $\textit{XMM-Newton}$ temperature data for $\delta_R > 0$ to obtain an SM $f_{gas}$ profile consistent with the universal gas mass fraction at the virial radius. The free parameters involved in the fitting procedure are reported in Table I, while $\delta_R$ and $l$ are fixed. The value of $\delta_R$ is changed in Eq. 1 until the SM gas mass fraction is consistent with the universal value at $R$. Instead, the 
$l$ value is fixed at 0.4 so as to have in the innermost regions $p_{nth} \simeq$ some percent of $p_{tot}$, in agreement with the \textit{Hitomi} observations of the Perseus cluster (higher $l$ values give higher levels of turbulence in the innermost cluster regions; the $l$ value does not enter in the calculation of $M_{tot}(R)$ (see Eq. A1) and has a negligible effect on the fitting of the temperature data). The agreement is reached for \textbf{$\delta_R=1.0\pm 0.1$} that implies $p_{nth} \simeq 50\% p_{tot}$ at the virial radius.
With these values of $\delta_R$ and $l$, 
$p_{nth}$ is $\simeq 17\%$ of the total pressure $p_{tot}$ at $r_{500}$.
Table I reports the free parameters values for the SM fits to the X-ray temperature data with $\delta_R=0$ and $\delta_R=1$ (see Fig. 1). $k_BT_R$ is the ICM temperature at the virial radius and $c$ is the concentration parameter (see Cavaliere et al. 2009); see Sect. 2 for the meaning of the other free parameters. Recently, Pearce et al. (2019) have shown that the function $\delta(r)$ is less able to follow the median  profile $\alpha(r)=p_{nth}(r)/p_{tot}(r)$ (= $\delta(r)/(1+\delta(r)$) derived from their simulations. However, this disagreement regards a region
with $r\lesssim 0.35r_{500}$ where the non-thermal pressure is negligible with respect to the thermal one (see Fig.s 4 and 5).

In the next Section, this high level of the non-thermal pressure component will be discussed together with the other SM thermodynamic quantities. In particular, it will taken into account the possible presence of gas clumpiness that can overestimate the electron gas density profile mostly in the cluster outskirts.

%\begin{equation}
%\delta(r) = \delta_R\, e^{-(R-r)^2/l^2}
%\end{equation}
%which decays on the scale $l$ toward the inside from a round maximum \textbf{(see Cavaliere et al. 2011)}. \textbf{The virial radius $R$ is assumed to be $2r_{500}$}.

\begin{figure*}
\begin{center}
\parbox{16cm}{
\includegraphics[width=0.4\textwidth,height=0.3\textheight,angle=0]{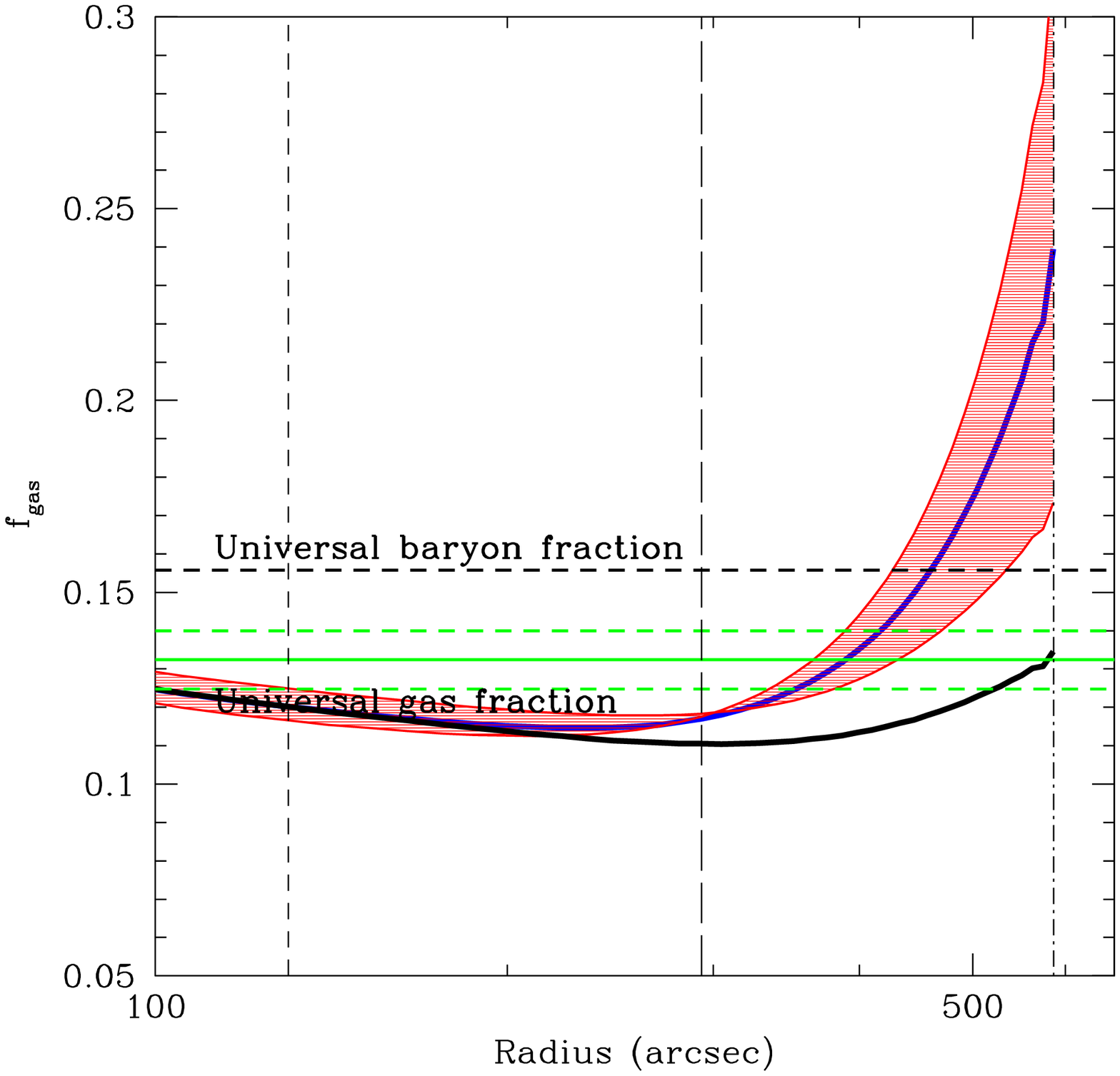}
\includegraphics[width=0.4\textwidth,height=0.3\textheight,angle=0]{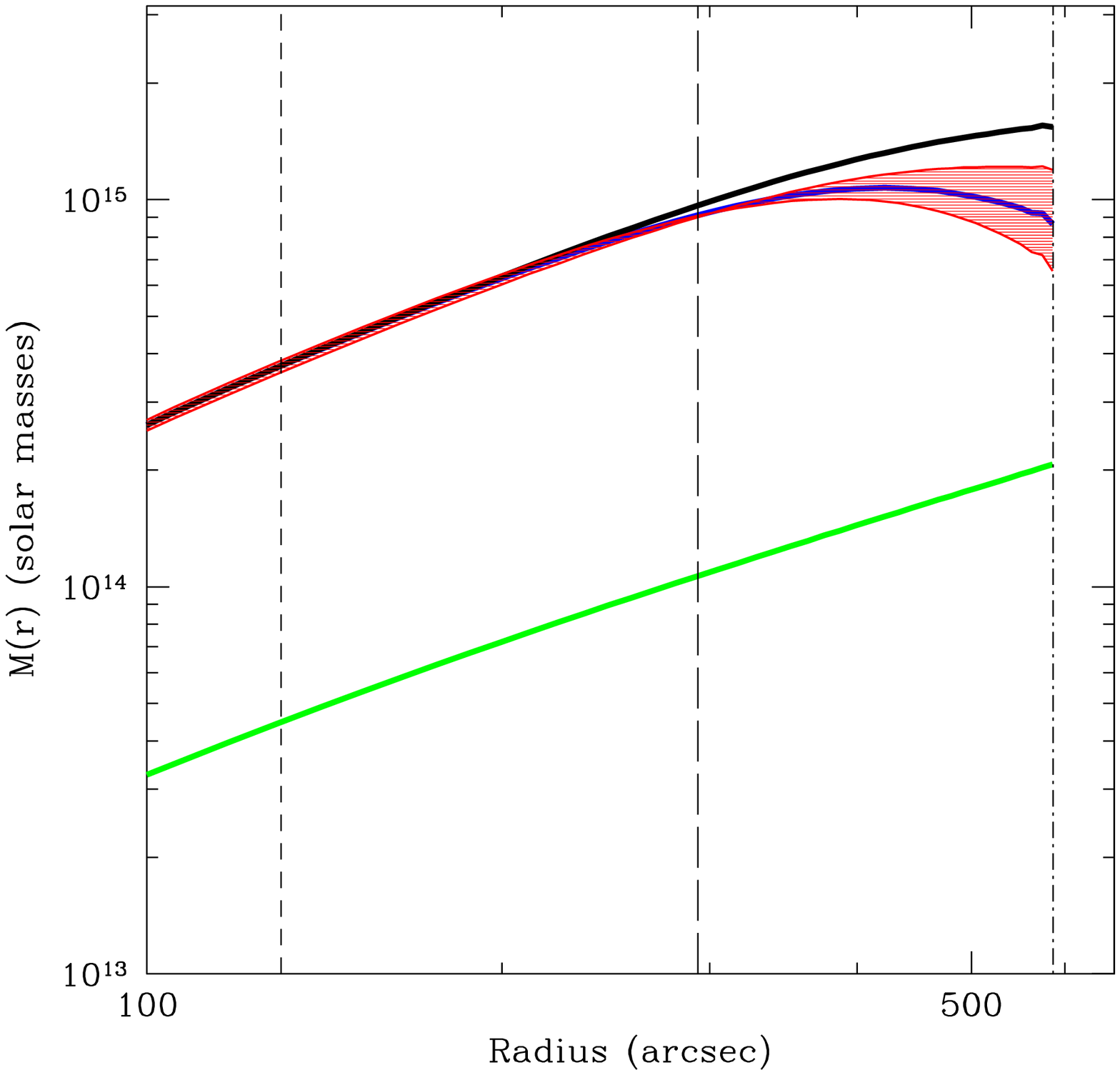}}
\parbox{16cm}{
\includegraphics[width=0.4\textwidth,height=0.3\textheight,angle=0]{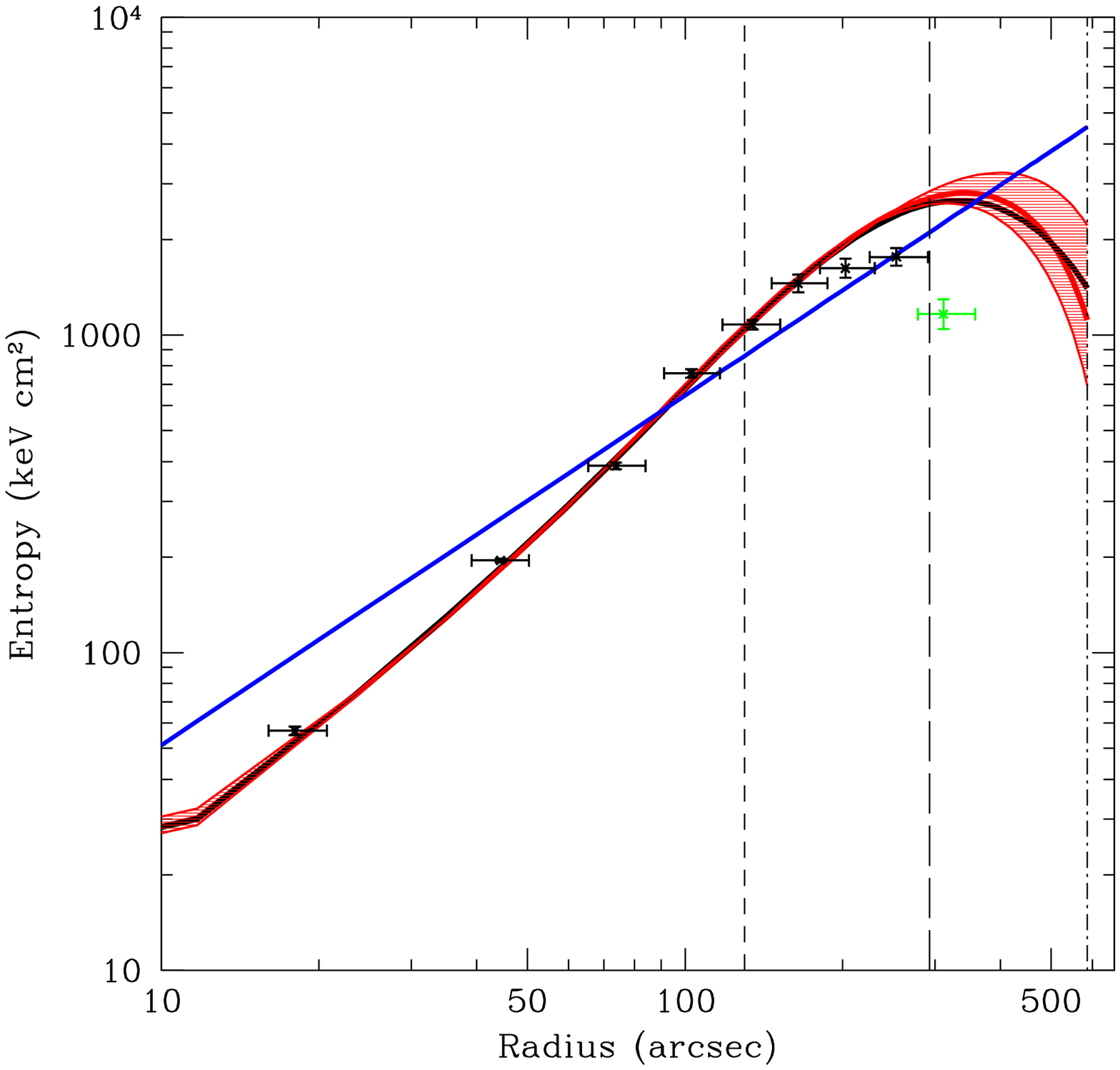}
\includegraphics[width=0.4\textwidth,height=0.3\textheight,angle=0]{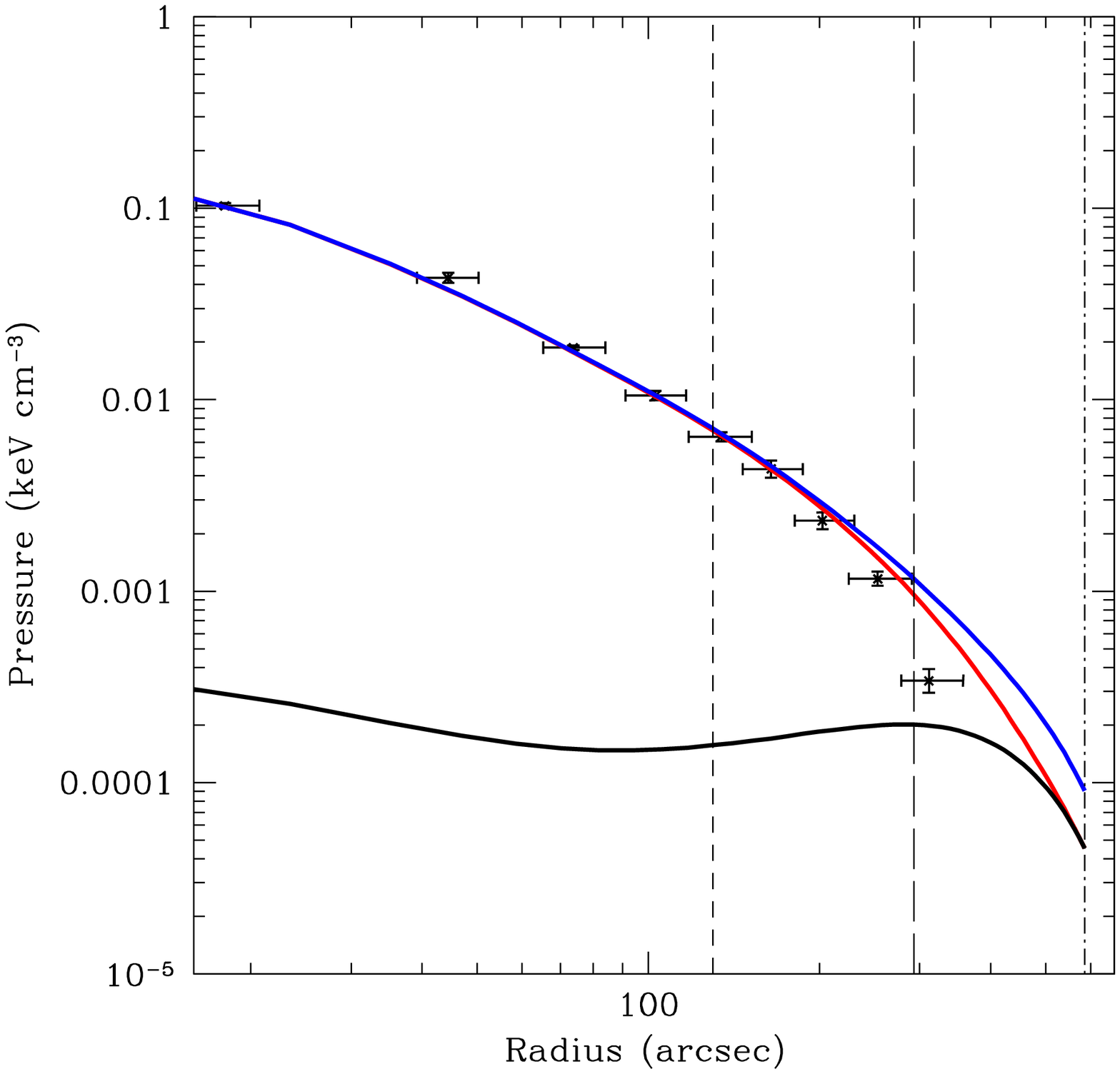}}
\caption{\textit{Top left panel}: Gas mass fraction $f_{gas}=M_{gas}/M_{tot}$ inferred from the SM temperature profile ($T_{SM}$), obtained from the fit to the $\textit{XMM-Newton}$ temperature profile ($T_X$, Romero et al 2019), and the gas density profile of Fig. 3 (continuous curve). The dashed area is obtained with $\delta_R = 0$ in Eq. A1, while the black line is derived with
$\delta_R = 1$ ($\alpha=p_{nth}/p_{tot}=50\%$) and $l = 0.4$. The dashed horizontal line represents the universal baryon fraction from \textsl{Planck} (Planck Collaboration XIII 2016), whereas the thick green line with the dashed green lines is the expected gas fraction corrected for the baryon fraction in the form of stars (Gonzalez, Zaritsky \& Zabludoff 2007). \textit{Top right panel}: The dashed area represents the total cluster mass derived with the above temperature and density profiles and with $\delta_R=0$, while the black line is for $\delta_R=1$ and $l=0.4$. The green curve is the gas mass derived with the continuous gas density profile of Fig. 3.
\textit{Bottom left panel}: The entropy points are derived from the X-ray temperature points of Fig. 1 and the continuous density curve
of Fig. 3 trough the relation $K=k_BT_X/n_e^{2/3}$. The dashed area is obtained from the SM fit to the temperature data (see Fig. 1) with 
$\delta_R=0$ and from the electron density of Fig. 3 (continuous curve). The black line represents the entropy profile for $\delta_R=1$ and $l=0.4$; the blue straight line is the expected power law increase with slope 1.1 (Voit 2005). \textit{Bottom right panel}: The pressure points are derived from the \textit{XMM-Newton} observations (Romero et al. 2019). The red curve is the thermal pressure derived
from the SM temperature profile with $\delta_R=1$ ($l=0.4$), and the density profile of Fig. 3 (continuous line) through the relation $p_{th},_{SM}=k_BT_{SM}n_e$. The black curve is the non-thermal pressure ($\alpha(R)\simeq 50\%$) and the blue curve is the total pressure $p_{tot}=p_{th}+p_{nth}$. In all panels, the thin lines represent the $1\sigma$ error, the vertical dashed, long-dashed and dot-dashed lines represent $r_{2500}$, $r_{500}$, and the virial radius $R$, respectively.}
\label{fig:corr}
\end{center}
\end{figure*}

\begin{figure*}
\begin{center}
\parbox{16cm}{
\includegraphics[width=0.4\textwidth,height=0.3\textheight,angle=0]{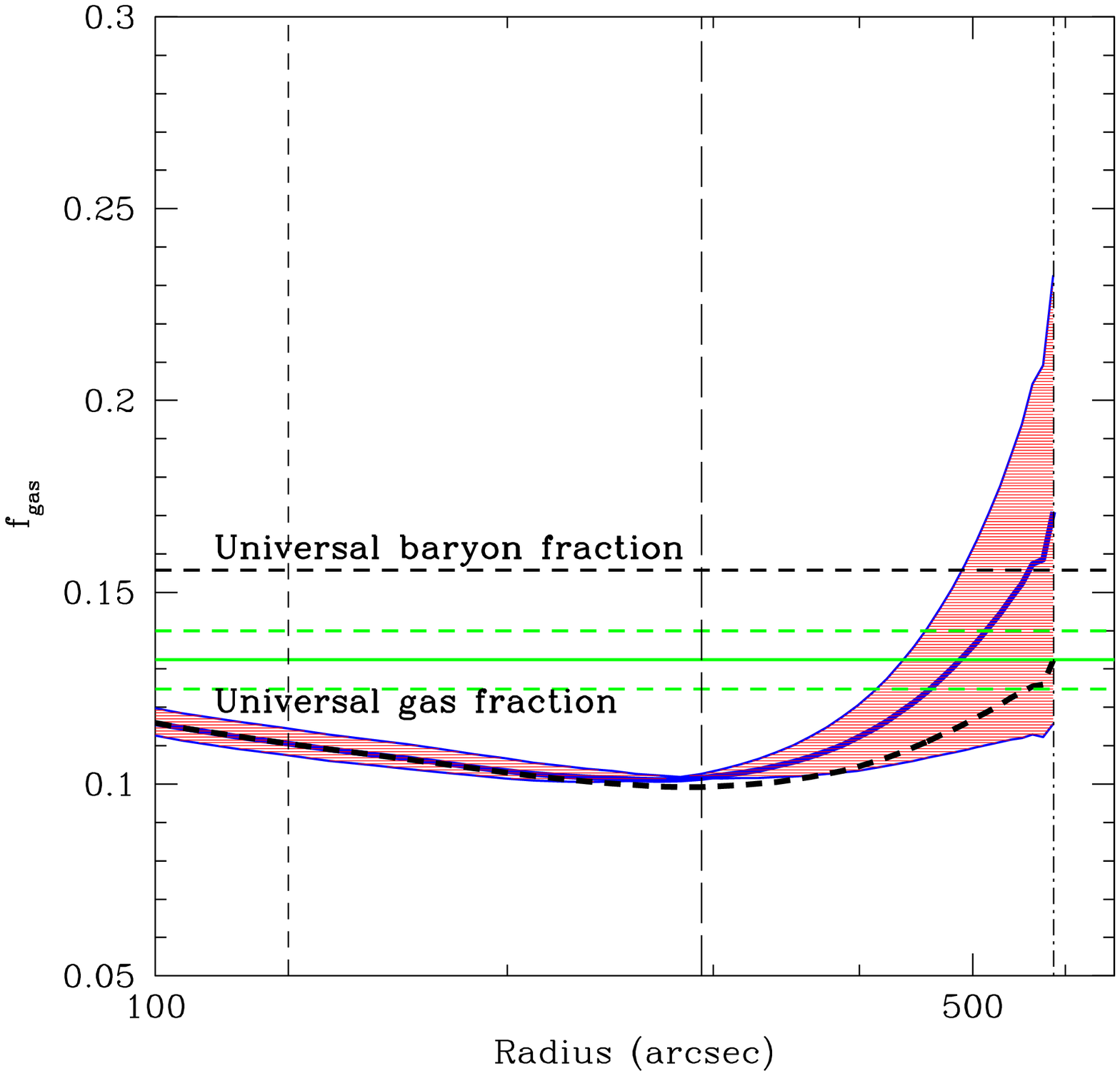}
\includegraphics[width=0.4\textwidth,height=0.3\textheight,angle=0]{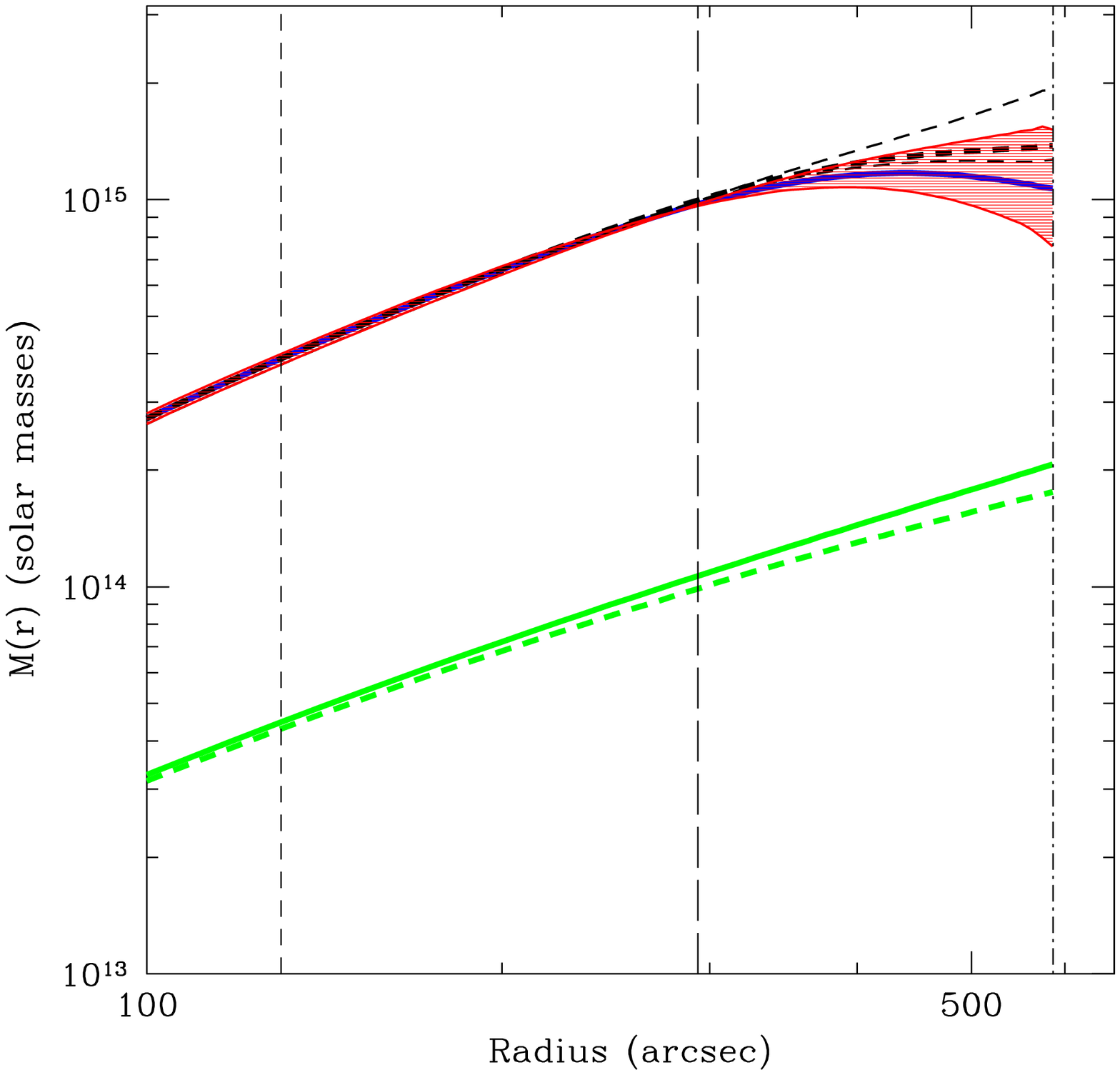}}
\parbox{16cm}{
\includegraphics[width=0.4\textwidth,height=0.3\textheight,angle=0]{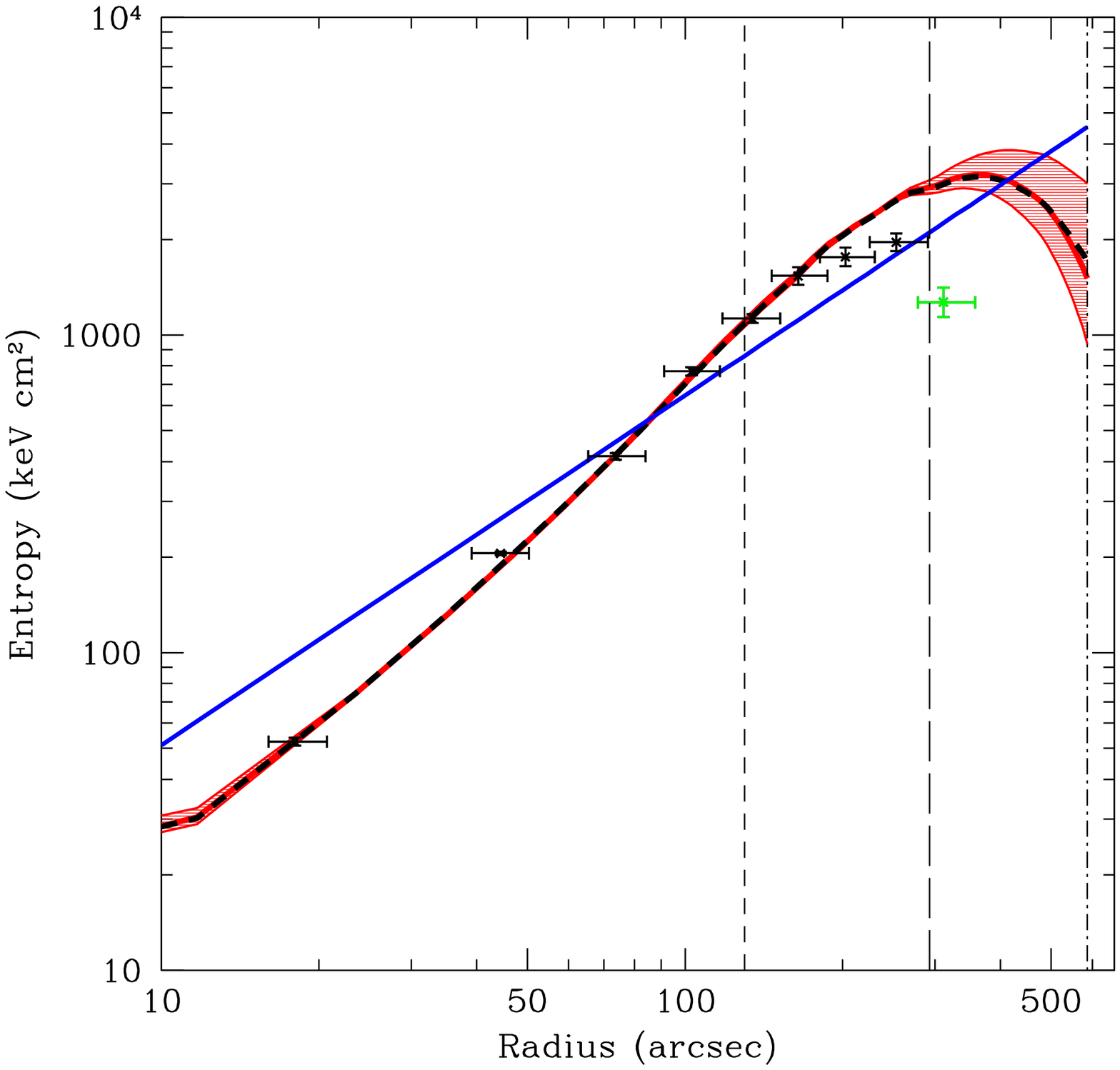}
\includegraphics[width=0.4\textwidth,height=0.3\textheight,angle=0]{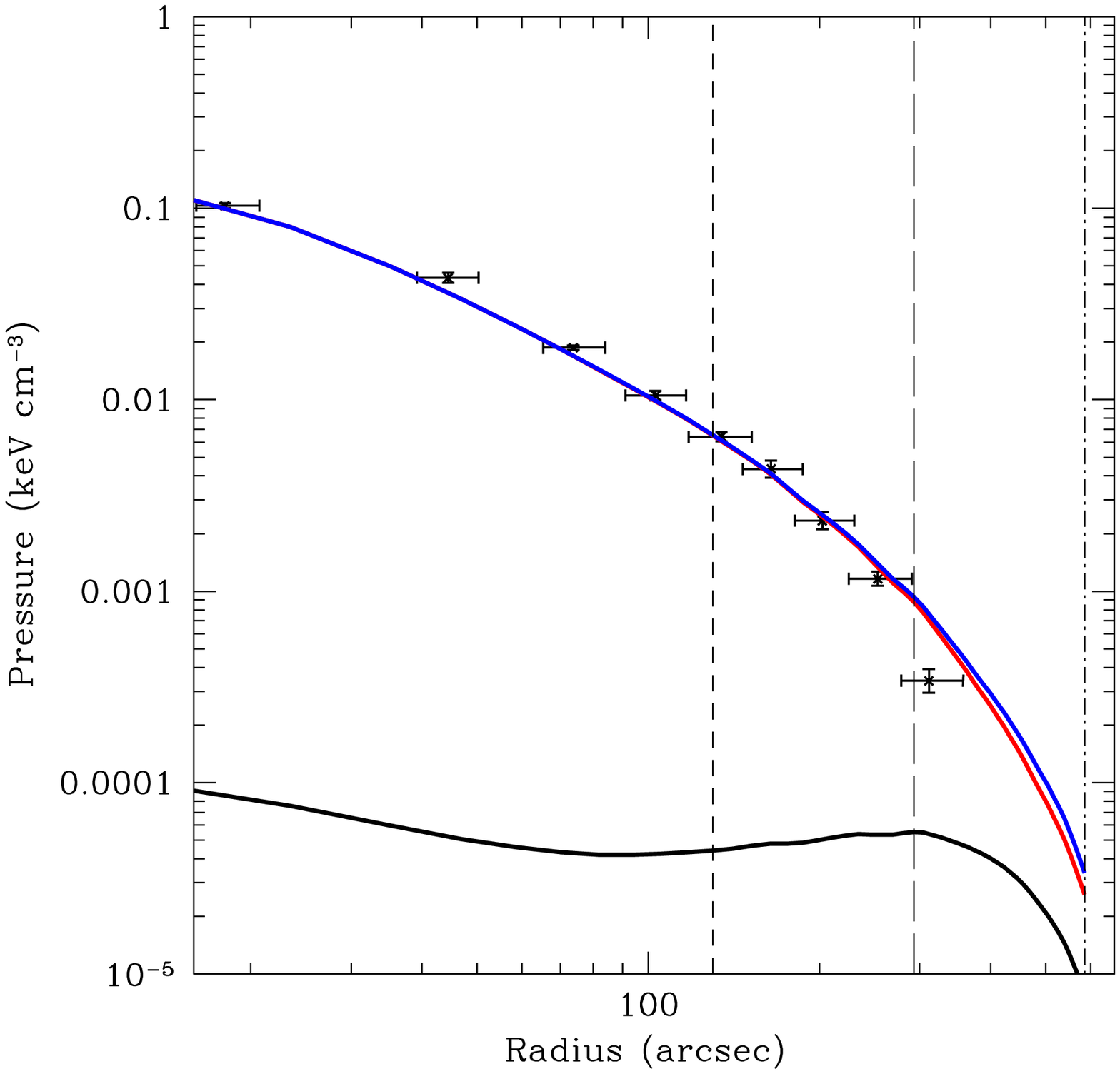}}
\caption{\textit{Top left panel}: Gas mass fraction $f_{gas}=M_{gas}/M_{tot}$ inferred from the SM temperature profile ($T_{SM}$), obtained from the fit to the $\textit{XMM-Newton}$ temperature profile ($T_X$, Romero et al 2019), and from the gas density profile of Fig. 3 (dashed curve). The dashed area is obtained with $\delta_R = 0$ in Eq. 1, while the dashed black line is derived with
$\delta_R = 0.3$ and $l = 0.4$ ($\alpha=p_{nth}/p_{tot}=20\%$). The dashed horizontal line and the thick green line as in Fig. 4. \textit{Top right panel}: The dashed area
represents the cluster total mass derived with the SM temperature profile ($\delta_R=0)$ and the dashed density profile of Fig. 3, while the dashed black line is derived 
for $\delta_R=0.3$ and $l=0.4$. The green curve is the gas mass derived with the continuous gas density of Fig. 3, while the dashed curve is the gas mass obtained with the gas density profile corrected for clumpiness.
\textit{Bottom left panel}: The entropy points are derived from the X-ray temperature points of Fig. 1 and the dashed density curve
of Fig. 3 through the relation $K=k_BT_X/n_e^{2/3}$. The dashed area is obtained from the SM fit to the temperature data with ($\delta_R=0$) and from the electron density corrected for clumpiness. The dashed black line represents the entropy profile for $\delta_R=0.3$ and $l=0.4$. The blue straight line as in Fig. 4. \textit{Bottom right panel}: The pressure points are derived from the \textit{XMM-Newton} temperature observations (Romero et al. 2019) and from the dashed density profile of Fig. 3. The red curve is the thermal pressure derived
from the SM temperature data with $\delta_R=0.3$ ($l=0.4$) and the density profile corrected for clumpiness through the relation $p_{th},_{SM}=k_BT_{SM}n_e$. The black curve is the non-thermal pressure ($\alpha(R)\simeq 20\%$) and the blue curve is the total pressure $p_{tot}=p_{th}+p_{nth}$. In all panels, 
the thin lines represent the $1\sigma$ error, the vertical dashed, long-dashed and dot-dashed lines represent $r_{2500}$, $r_{500}$, and the virial radius $R$, respectively.}
\label{fig:corr}
\end{center}
\end{figure*}

\begin{table}
	\centering
	\caption{Best fit parameters of the SM analyses.}
	\label{tab:SM_table}
	\begin{tabular}{lccc} % four columns, alignment for each
		\hline
		& $\delta_R=0$  &  & $\delta_R=1$ \\
		\hline
		$k_BT_R  \rm (keV)$  & $1.127^{+1.108}_{-0.499}$ &  & $1.417^{+1.387}_{-0.517}$ \\
		$k_c/k_R$ & $(7.2\pm 1.7)\times 10^{-3}$ &  & $(6.3\pm 1.6)\times 10^{-3}$ \\
		$a$  & $1.99\pm 0.64$ &  & $2.07\pm 0.81$ \\
		$c$ & $4.54\pm 1.51$ &  & $3.97\pm 1.23$ \\
		$r_b/R$ & $0.16\pm 0.08$ &  & $0.16\pm 0.09$ \\
		$a^{\prime}$ & $0.99\pm 0.58$ &  & $0.99\pm 0.57$ \\
		\hline
	\end{tabular}
\end{table}

\section{Discussion and Conclusions}

In this paper, the SM analysis is centered on the gas temperature and density data reported by \textit{XMM-Newton} observations of the galaxy cluster Zwicky 3146. The SM fit to the temperature profile is obtained assuming a modified entropy distribution that flattens at distances greater than $r_b$ (see Sect. 2 and Table I), as reported by $\textit{Suzaku}$ observations in several clusters. It must be noted that the SuperModel is able to evidence an entropy flattening in the cluster outskirts fitting temperature data limited to $\sim r_{500}$, as already shown for the stacked temperature profile of the X-COP clusters (Fusco-Femiano 2019). The SM temperature profile 
is in agreement with that derived from the gNFW model, while the temperature profile from the NP model
results steeper in the cluster outskirts, the latter two profiles come from the SZ pressure data. 
The gas mass fraction, obtained from the SM temperature profile and from the gas density profile shown in Fig. 3 (continuous line), compared with
the universal value indicates the presence of a non-thermal pressure support in the cluster outskirts.
Neglecting this component in the HE equation implies a nonphysical behaviour of the total mass (see Fig. 4) and biases in the physical understanding of the thermodynamic quantities. 

The level at the virial radius of the ratio $\delta_R=p_{nth}/p_{th}$ (and of $\alpha(R)=p_{nth}/p_{tot}=\delta_R/(1+\delta_R$)) is estimated
using the function $\delta(r)$. The SM $f_{gas}$ profile obtained with $\delta_R = 0$ results above the universal value at the virial radius. The agreement is reached for $\delta_R\simeq 1$ which involves a non-thermal pressure component $\simeq 50\%$ of the total pressure. 

Hydrodynamic simulations show that $p_{nth}$ due to turbulent gas motions within the intracluster medium increases from $\approx 10\%$ at $r_{500}$ to $\approx 30\%$ at the virial radius (Lau et al. 2009; Vazza et al. 2009; Battaglia et al. 2012; Pearce et al. 2019). Recently, Vazza et al. (2018) estimate a support of $\sim 10\%$ at $r_{200}$ constraining which fraction of the gas kinetic energy effectively provides pressure support in the cluster's gravitational potential. 

The high value of the ratio $\alpha(R)\simeq 50\%$, obtained from the SM analysis, with respect to the values found in numerical simulations, leads to suppose that this discrepancy may be attributed to the presence in the gas density profile of accreting substructures and inhomogeneities. In this case the gas density inferred from the X-ray surface brightness is overestimated by a factor $\sqrt(C)$ where $C \equiv \langle n^2_{gas}\rangle / \langle n_{gas}\rangle^2$ is the clumping factor. Nagai \& Lau (2011) have derived the median profile of $C$ for 
clusters with $T_{gas} > 10^6 K$ and $M_{200}\gtrless 10^{14} h^{-1} M_{\odot}$. The median profile $\sqrt(C)$ for $M_{200} > 10^{14} h^{-1} M_{\odot}$ has been applied to the electron gas density obtained from the $\textit{XMM-Newton}$ observations, yielding the dashed curve of Fig. 3.

A new SM analysis has been carried out using the gas density profile corrected for clumpiness. The presence of inhomogeneities is expected to be irrelevant for the X-ray temperature data at $r\lesssim r_{500}$ (see Ghirardini et al. 2019). This is confirmed by the last two X-ray points at $r\sim r_{500}$ of the stacked temperature profile of the X-COP cluster sample that are consistent with the SZ temperature points less affected by clumping (Fusco-Femiano 2019). The universal gas mass fraction is now satisfied by the SM $f_{gas}$ profile for $\alpha(R) \simeq 20\%$ (or $\delta_R\simeq 0.3\pm 0.1$; $l$ is fixed at 0.4) consistent with the values reported by numerical simulations. With these values of $\delta_R$ and $l$ $p_{nth}$ is $\simeq 6\%p_{tot}$ at $r_{500}$.
The new $f_{gas}$, mass, entropy and pressure profiles are shown in Fig. 5. It must be noted that the
the universal value of the gas mass fraction is within the $1\sigma$ error of the SM $f_{gas}$ profile derived for $\delta_R=0$ and with the corrected gas density profile. However, the decreasing total mass profile in the cluster outskirts points out the presence of a non-thermal pressure support.

Regarding the cluster mass, the SM analysis based on the the X-ray observations reports values of $M_{500}$, derived from the hydrostatic equilibrium method, consistent with that reported by R19 based on the SZ data ($8.29^{+1.93}_{-1.24}(stat)^{+0.74}_{-0.54}(sys)\times 10^{14}M_{\odot}$). In fact, using the gas density not corrected for clumpiness and a non-thermal pressure 
component of $\simeq 50\%$ of the total pressure at the virial radius, the SM analysis yields $M_{500}=9.65\pm 0.3\times 10^{14}M_{\odot}$. With the gas density corrected for clumpiness and $\alpha(R)\simeq 20\%$, $M_{500}$ is $9.95\pm 0.16\times 10^{14}M_{\odot}$. An evident tension is between these estimates and
the value of $M_{500}=3.88^{+0.66}_{-0.58}\times 10^{14}M_{\odot}$ derived from weak lensing by Okabe \& Smith (2016).
A total mass $M_R=1.53^{+0.55}_{-0.35}\times 10^{15}M_{\odot}$ is predicted from the first SM analysis and $M_R=1.38^{+0.55}_{-0.11}\times 10^{15}M_{\odot}$ from the second. Not taking into account the non-thermal pressure in the HE underestimates $M_{500}$ of only some percent (3-6)\% for the two different gas density profiles. Instead,
at the virial radius, the analysis with the gas density not corrected for clumpiness gives an underestimation of $\sim 40\%$, while is $\sim 20\%$ using the gas density corrected for clumpiness. The last value is in the range (10-30)\% usually reported in literature (see Biffi et al. 2016; Khatri \& Gaspari 2016: Hurier \& Angulo 2018; Ettori et al. 2019).

The overestimation of the gas density profile caused by clumping determines an underestimation of the entropy profile by a factor $C(r)^{1/3}$. 
Considering that the $n_e$ profile is now corrected for clumpiness, the entropy flattening starting at $r\gtrsim r_{500}$ (see Fig. 5) is due to the steep decline of the temperature, in agreement with the conclusions of Okabe et al. (2014) in their joint X-ray and weak lensing study of four relaxed galaxy clusters observed by \textit{Suzaku} and \textit{Subaru} out to the virial radii. The same conclusion is found by the SM analysis of the X-COP cluster sample. 
The SM entropy profiles reported in Fig.s 4 and 5 result steeper, in the region  $23^{\prime\prime}-138^{\prime\prime}$ (100-600 kpc), of the power law 
with slope 1.1 expected under pure gravitational collapse (Voit 2005). The slope from the SM analysis of the X-ray data is $\sim 1.56$
not much different from the value of 1.44 derived with the gNFW model (R19) applied to the SZ data. A start of an entropy flattening toward the core is at $r\lesssim 20^{\prime\prime}$.

In conclusion, the SM analysis of the galaxy cluster Zwicky 3146, based on the X-ray temperature data limited to $r_{500}$, predicts that the properties reported by \textit{Suzaku} observations in several relaxed clusters could be present also in the outskirts of Zwicky 3146. Namely, a steep temperature decline, the flattening of the entropy profile, and a decreasing mass profile due to the break of the hydrostatic equilibrium for the presence of turbulence in the cluster outskirts. The SM analysis of Zwicky 3146 appears to be a further confirmation (see the SM analysis of the X-COP clusters, Fusco-Femiano 2019) that the temperature profiles observed by \textit{XMM-Newton}
are consistent with the rapid decline of the temperature in the cluster outskirts reported by the \textit{Suzaku} observations. Besides, the SM analysis highlights the possible presence of clumpiness in the gas density profile considering the derived high level of the ratio $\alpha(R)$  
with respect to the values found in numerical simulations. However, also considering the clumpiness, turbulence and entropy flattening are still present in the cluster outskirts. As suggested by Lapi, Fusco-Femiano \& Cavaliere (2010), the turbulence may be related to the weakening of the accretion shocks in relaxed clusters which induces an increase of the bulk inflow energy
in the cluster outskirts and a reduction of the thermal energy. This leads to the saturation of the entropy production determining the observed rapid decline of the temperature. This scenario
is in agreement with the significant azimuthal variations of the electron density, temperature, and entropy reported by $\textit{Suzaku}$
observations in some clusters. In fact, the weakening degree of the accretion shocks may be more pronounced in cluster sectors adjacent to low density regions of the 
surrounding environment.

\section*{Acknowledgements}
I thank the referee for valuable comments.
%Work partially supported by PRIN MIUR 2015 `Cosmology and Fundamental Physics: illuminating the Dark Universe with Euclid'.

\appendix

\section{Hydrostatic cluster mass in presence of turbulence}

%As suggested by Cavaliere et al. (2011), the weakening of the accretion shocks involves more
%bulk energy to flow through the cluster, and drive turbulence into the
%outskirts. Turbulent motions originate at the virial
%boundary and then they fragment downstream into a dispersive
%cascade to sizes $l$. Numerical simulations report a negligible contribution of the
%turbulent energy in the cores of relaxed clusters and an increase into the outskirts of the ratio
%of the turbulent to thermal energy  (e.g., Vazza et al. 2011). For $\delta(r)$ the functional shape
%\begin{equation}
%\delta(r) = \delta_R\, e^{-(R-r)^2/l^2}
%\end{equation}
%which decays on the scale $l$ inward of a round maximum is used. This profile of
%$\delta(r)$ concur with the indication of numerical simulations (Lau et al.
%2009; Vazza et al. 2011).

Here it is reported the equation to derive from the hydrostatic equilibrium the total cluster mass $M_{tot}$ within $r$ in presence of an additional non-thermal pressure component in the HE equation (Fusco-Femiano \& Lapi 2013)

\begin{equation*}
M_{tot}(r) = - \frac{k_B [T(r)(1 +\delta(r)] r^2 }{\mu m_p G}\,\times
\end{equation*}
\begin{equation*}
\left\{\frac{1}{n_e(r)}\frac{d n_e(r)}{d r} +
\frac{1}{T(r)[(1+\delta(r)]}\frac{d T(r)[1 +
\delta(r)]}{d r}\right\}
\end{equation*}
\begin{equation*}
= - \frac{k_B [T(r)(1 +\delta(r)] r^2}{\mu m_p G}\,\times
\end{equation*}
\begin{equation}
\left[\frac{1}{n_e(r)}\frac{d n_e(r)}{d r} +
\frac{1}{T(r)}\frac{d T(r)}{d r} + \frac{\delta(r)}{1
+ \delta(r)} \frac{2}{l^2}(R - r)\right]~.
\end{equation}
\par\noindent
where $\delta(r)=p_{nth}/p_{th}=\delta_R\, e^{-(R-r)^2/l^2}$

\par\noindent
The gas mass is given by
$$M_{gas} = 4\pi \mu_e m_p\int{\rm d}r~{n_e(r) r^2}$$
where $\mu_e \sim 1.16$ is the mean molecular weight of the electrons.

% Don't change these lines
\bsp	% typesetting comment
\label{lastpage}
\end{document}